 \documentclass[preprint2,longabstract]{aastex}

\slugcomment{revision - \today}

\shorttitle{heat flux transition}
\shortauthors{Bale et al.}

\begin{document}

\title{Electron heat conduction in the solar wind: transition from Spitzer-H\"{a}rm to the collisionless limit }

\author{S. D. Bale\altaffilmark{1, 2}, M. Pulupa\altaffilmark{2}, C. Salem\altaffilmark{2}, C. H. K. Chen\altaffilmark{2}, and E. Quataert\altaffilmark{3, 1}}

\altaffiltext{1}{Physics Department, University of California, Berkeley.}
\altaffiltext{2}{Space Sciences Laboratory, University of California, Berkeley.}
\altaffiltext{3}{Astronomy Department, University of California, Berkeley.}

\begin{abstract}
We use a statistically significant set of measurements to show that the 
field-aligned electron heat flux $q_\parallel$ in the solar wind at 1 AU is consistent with the Spitzer-H\"{a}rm 
collisional heat flux $q_{sh}$ for temperature gradient scales larger than a few mean free 
paths $L_T \gtrsim 3.5 ~\lambda_{fp}$.  This represents about 65\% of the measured data  
and corresponds primarily to high $\beta$, weakly collisional plasma ('slow solar wind').  
In the more collisionless regime $\lambda_{fp}/L_T \gtrsim 0.28$, the electron heat flux is 
limited to $q_\parallel/q_0 \sim 0.3$, independent of mean free path, where $q_0$ is the 'free-streaming' value; the measured $q_\parallel$ does not 
achieve the full $q_0$. 
This constraint  $q_\parallel/q_0 \sim 0.3$ 
might be attributed to wave-particle interactions, effects of an interplanetary electric potential, or inherent flux limitation.  We also show a $\beta_e$ dependence to these results 
that is consistent with a local radial electron temperature profile $T_e \sim r^{-\alpha}$ 
that is a function of the thermal electron beta $\alpha = \alpha(\beta_e)$ and that the $\beta$ dependence 
of the collisionless regulation constraint is not obviously consistent with a whistler heat flux instability.  It may be that 
the observed saturation of the measured heat flux is a simply a feature of collisional transport.  We discuss the 
results in a broader astrophysical context.
\end{abstract}


\section{Introduction}

Thermal conduction in the solar wind provides an important mode of energy transport and determines in part  
the radial electron temperature profile.  The conductive, magnetic field-aligned electron heat flux is 
defined as $q_{\parallel} = -\kappa_\parallel \nabla_\parallel T_e$ where $\kappa_\parallel$ is the thermal 
conductivity coefficient and $T_e (r)$ is the electron temperature.  In a fully collisional plasma, the Spitzer-H\"{a}rm thermal conductivity \citep{Spitzer++53} is 
$\kappa_\parallel = \kappa_{SH} \sim 3.16 ~n_e T_e \tau_{e}/m_e$.  Spitzer-H\"{a}rm (SH) theory assumes that the electron distribution function $f(\vec{v})$ remains 
approximately Maxwellian as it evolves through a temperature gradient scale, which corresponds to assuming that the Knudsen number $K \sim \lambda_{fp}/L_T$ is a small parameter , 
where $L_T = T_e/|\partial T_e/\partial r|$ = $R/\alpha$ for $T(r) \sim r^{-\alpha}$, where R = 1 AU, and $\lambda_{fp} = v_e \tau_{e}$ is the mean free path.  In particular, 
SH theory assumes that the distribution function $f \approx f_0 + f_1$, with $f_0$ an isotropic Maxwellian and $f_1$ an anisotropic ($\cos \theta$) perturbation, such that 
$f_1 \leq f_0$ and dropping $\mathcal{O}(f_1^2)$ terms \citep{Spitzer++53}.
The electron collision time $\tau_{e}$ goes like $T_e^{3/2}/n_e$, therefore the 
thermal conductivity scales like $\kappa_{SH} \propto T_e^{5/2}$.  Thus to maintain a constant conductive luminosity $L = 4 \pi r^2 q_\parallel$, 
the wind must have $T_e(r) \sim r^{-2/7}$.  Measurements of the solar wind electron temperature profile generally show 
a power law profile $T_e(r) \sim r^{-\alpha}$ with values of $\alpha$ from 0.2 to 0.7 \citep{Marsch++89}, not 
inconsistent with $\alpha = 2/7 \approx 0.286.$  Of course, energy input from turbulent dissipation is likely to be important as well.

The solar wind electron population at 1 AU consists primarily of a cool  Maxwellian 'core' ($\sim$10 eV, $\sim$95\% density), a suprathermal  'halo' ($\sim$70 eV, $\sim$4\% density), and an antisunward 'strahl' 
population ($\sim$100-1000 eV, $\sim$1\% density).  The core is nearly isotropic, while the halo and strahl exhibit clear temperature anisotropies.  
Since the electrons are subsonic ($v_e\gg v_{sw}$), the maximum available heat flux corresponds to transport of the 
full thermal energy $3/2 ~n_e k_b T_e$ at the thermal speed $v_e$; this is often called the 'free-streaming' or saturation 
heat flux $q_0 = 3/2 ~n_e k_\mathrm{B} T_e v_e$ \citep{Parker64, Roxburgh74, Cowie++77}.  The
ratio $q_{sh}/q_0$ is proportional to the Knudsen number $K \sim \lambda_{fp}/L_T$, the small parameter in SH theory \citep{Cowie++77, Salem++03},

\begin{equation}
\frac{q_{sh}}{q_0} = \frac{-\kappa_{SH} \nabla_\parallel T_e}{(3/2) n ~k_b T_e v_e} \approx 1.07 ~\frac{\lambda_{fp}}{L_T}
\end{equation}

and this scaling of $q_\parallel/q_0$ with  $\lambda_{fp}/{L_T}$ provides a clear test of SH theory.  
Assuming a value of $\alpha$ in $L_T$, we can compute the measured $q_{||}/q_0$ vs $\lambda_{fp}/L_T$ and compare to Equation (1)
for $q_{sh}/q_0$; 
the only free parameter is $\alpha$, the temperature gradient exponent.

In this short paper we show that the measured $q_\parallel$ in the solar wind is consistent with $q_{sh}$ until  $\lambda_{fp}/L_T$ becomes 
as large at 0.3.  Beyond that, in the collisionless regime, we find that $q_\parallel \sim 0.3 ~q_0$ independent of mean free path.  Collisionless 
regulation of heat flux has been discussed in the context of wave-particle interactions \citep{Gary++94} and escape from an 
interplanetary electric potential \citep{Perkins73, Hollweg74}.  We divide our data into intervals of electron thermal $\beta_e$ (ratio of electron
thermal pressure to magnetic field pressure) and compare to theoretical threshold values for whistler and magnetosonic instability constraints on 
heat flux \citep{Gary++94}.  We find that the whistler instability overconstrains the measurements while the magnetosonic instability is more 
consistent.  We also find that the data fit better to the SH relationship (in the collisional regime) if there is a $\beta_e$ 
dependence to the temperature profile index $\alpha$.  This may reflect the $\beta_e$ dependence of a wave heating mechanism, or may be 
a proxy for another parameter, such as collisional age.

\section{Measurements}

We use measurements of the solar wind electron distribution function from the Three Dimensional 
Plasma (3DP) instrument \citep{Lin++95, Pulupa++13} on the NASA Wind spacecraft.  The 3DP instrument 
uses two separate sensors - EESA-L and EESA-H - to measure the full 3D distribution 
function $f(\vec{v})$ in 88 angular bins from $\sim$1 eV to $\sim$30 keV, once per spacecraft spin (3 s).  Each EESA sensor is 
a 'top hat' electrostatic analyzer \citep{Carlson++82} designed to measure solar wind 
thermal electrons (EESA-L) and suprathermals (EESA-H) each in 15 log-spaced energy steps (few eV to 1.1 keV for EESA-L and 
100 eV to 30 keV for EESA-H).  Measurements from the 
two detectors are combined to form a single distribution function and this distribution function is 
corrected for spacecraft floating potential effects using quasi-thermal noise measurements as an 
absolute plasma density benchmark; low energy monopole corrections (few Volts) are less 
important for higher-order moments such as heat flux, however dipole fields may introduce errors of 5\% to the heat flux 
moment \citep{Pulupa++13}.
We use 155,182 independent measurements of $f(\vec{v})$ from two, 2-year intervals:
1995-1997 (solar minimum) and 2001-2002 (solar maximum) and include only 'ambient' solar 
wind (no CMES, foreshock, etc.).  Intervals of 'bi-directional' heat flux (usually associated with CMEs) 
are also excluded.  We compute the electron heat flux moment as

\begin{equation}
\vec{q}_e = \frac{1}{2} \int dv^3 f(\vec{v}) w^2 \vec{w}
\end{equation}

from the measurements, where the secular velocity is $\vec{w} = \vec{v} - \vec{v}_{b}$ and $\vec{v}_b$ is the bulk speed.  Here we consider the (dominant) magnetic field-aligned component 
of the heat flux $q_{\parallel} = \vec{q} \cdot \hat{B}$.  The thermal electron properties $q_0$, $\lambda_{fp}$, and $\beta_e$ are computed 
using measurements of the 'core' electron population, a point that we discuss below.

Figure 1 shows the joint probability distribution of 
$q_\parallel/q_0$ and $\lambda_{fp}/L_T$ normalized to the peak value in each $\lambda_{fp}/L_T$ histogram and Equation (1) is 
over plotted as a diagonal line in the top panel.  The number of points in each $\lambda_{fp}/L_T$ bin is shown in the lower panel and 
a temperature exponent of $\alpha \sim 2/7$ is used to calculate $L_T$.  It is apparent that $q_\parallel/q_0$ tracks Equation (1) over much of its range.

\begin{figure}
\includegraphics[width=85mm,height=90mm, scale=1,clip=true, draft=false]{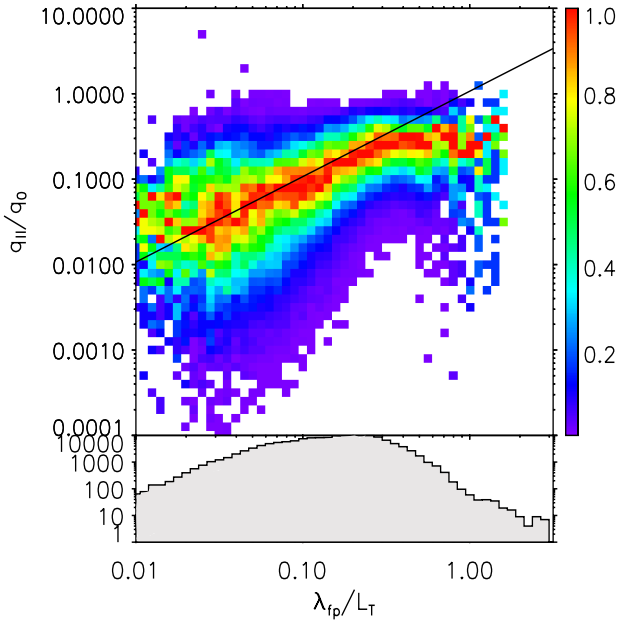}
\caption{Joint normalized distribution of the normalized electron heat flux $q_\parallel/q_0$ and 
the temperature Knudsen number $K  = \lambda_{fp}/L_T$ (top panel) and the distribution 
of data with $\lambda_{fp}/L_T$ (bottom panel). The diagonal line in the top panel is the Spitzer-H\"{a}rm relationship 
(Equation (1)).}
\end{figure}

Figure 2 shows the mode (most probable value) in each bin of $\lambda_{fp}/L_T$.  The modes are calculated directly from the distribution of data (red points) and 
from log-normal fits (black points).  In this Figure and in Figure 1, the Spitzer-H\"{a}rm relationship in Equation (1) appears to be a reasonably good approximation to the data.  
At about $\lambda_{fp} \simeq 0.28 ~L_T$, the measured data breaks from the SH line and flattens to a fixed value of $q_\parallel \sim 0.29~ q_0$.  Approximately 65\% of our measurements lie in the SH regime, the other 35\% in the collisionless regime $q_\parallel \simeq 0.29 ~q_0$ (bottom panel 
of Figure 2).  The so-called 'collisional age' $A = v_{sw} \tau$ is often used to measure collisional evolution \citep{Salem++03, Bale++09}, especially when considering 
interactions with ions convecting at the solar wind speed $v_{sw}$.  The collisional age correlates strongly with solar wind speed; fast solar wind is more collisionless (hot and rarified) and 
slow wind is more collisional (cooler and dense);  therefore data in Figure 1 and 2 in the collisional regime ($\lambda_{fp}/L_T \ll 1$) is primarily slow wind, while 
collisionless data ($\lambda_{fp}/L_T \gg 1$) is primarily fast wind.

\begin{figure}
\includegraphics[width=80mm,height=90mm, scale=1,clip=true, draft=false]{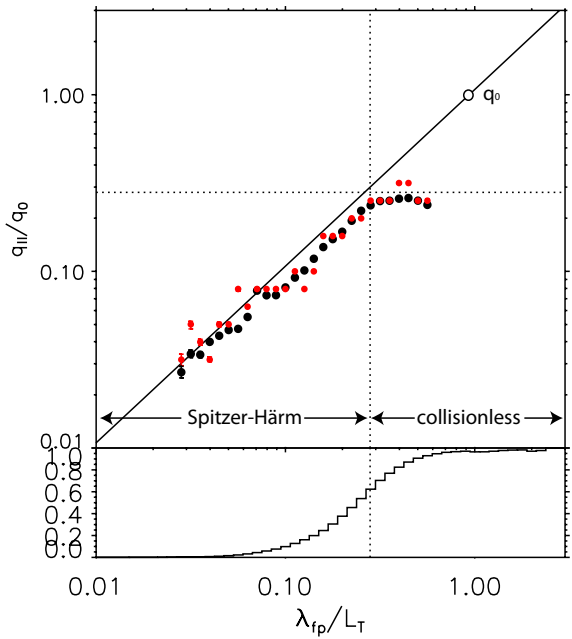}
\caption{Plot of the most probable values (modes) of $q_\parallel/q_0$ in bins of $\lambda_{fp}/L_T$ with standard 
error errorbars (very small).  Red points are the computed modes, black points are modes determined 
from log normal fits.  Again, the black diagonal line in the upper panel is the SH Equation (1).  The measured 
values depart from SH behavior above $\lambda_{fp}/L_T \sim q_\parallel/q_0 \gtrsim 0.28$.   The cumulative distribution in the bottom 
panel shows that approximately 65\% of this dataset corresponds to SH heat flux.  A symbol at $q_\parallel = q_0$ show the free-streaming value.}
\end{figure}

\section{Electron $\beta$ dependence}

\begin{figure}
\includegraphics[width=83mm,height=110mm, scale=1,clip=true, draft=false]{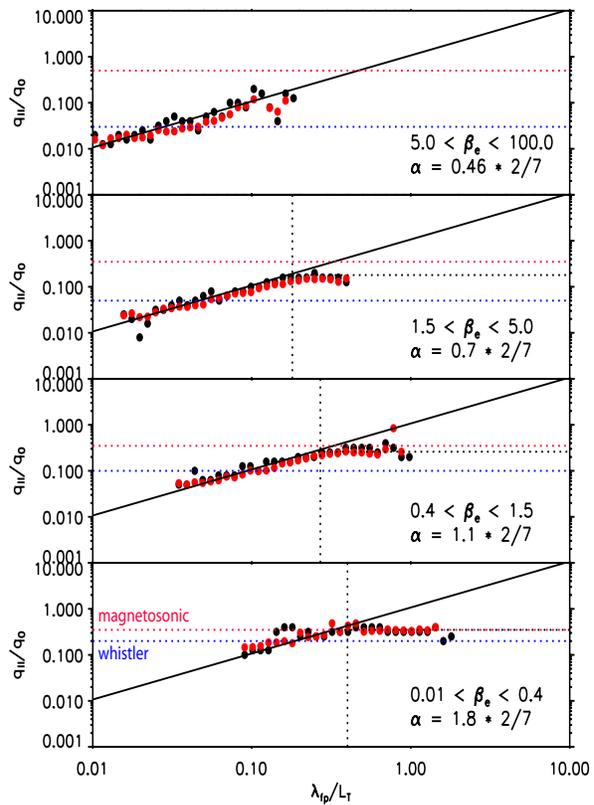}
\caption{Data from Figure 2 broken into four intervals of electron thermal $\beta_e$.  In each interval, the data are consistent with the SH 
relationship with the temperature gradient index
$\alpha$ as a free parameter.  The $\beta_e$ dependence of $\alpha$ is summarized in Table 1.  Dotted horizontal lines 
are the thresholds of the magnetosonic instability (red) and  whistler instability (blue), both from Gary et al (1994).  The whistler 
instability appears inconsistent with the heat flux levels in the collisionless regime.}
\end{figure}

In Figure 3, we break this data into 4 intervals of electron thermal beta $\beta_e = (n_e k_\mathrm{B} T_e)/(B^2/2 \mu_0)$.  Since $\lambda_{fp}/L_T \propto 1/n_e$ and 
density variations dominate the pressure variations, the high $\beta_e$ plasma corresponds to collisional plasma (small $\lambda_{fp}/L_T$).  This evolution 
can be seen in the panels of Figure 3, organized from high to low $\beta_e$ (top to bottom); the points move towards the right (towards the collisionless 
regime).  As the clusters of points move to the right, they maintain the SH-like power law behavior, but require different values of $\alpha$ (the temperature 
profile index) to conform to the curve - note that $\alpha$ is the only free parameter.  Alignment to the SH curve give corrected values and suggest a 
$\beta_e$ dependence to the electron temperature profile index $\alpha$ and to the breakpoint $\Lambda_b$ between the collisional SH and the collisionless 
heat flux regimes.  Table 1 summarizes these results.

Figure 3 also shows the $\beta_e$-dependent instability-limited electron heat flux values for both the magnetosonic and whistler instabilities as calculated by Gary et al. (1994).  
We use most probable values of $\beta_e$ in each interval and scale the corresponding ($\gamma_{max} = 10^{-3} ~\Omega_p$) threshold from Figure 1 of Gary et al. (1994): magnetosonic-limited 
values are shown as horizontal dotted red lines, while whistler instability-limited heat flux is shown as blue dotted lines.  It can be seen that the heat flux-driven whistler 
instability overconstrains the data; the magnetosonic instability underconstrains the data in the SH regime, but is in fact consistent with the limiting electron heat flux 
in the collisionless regime.

\begin{table}[htdp]
\begin{center}
ÊÊÊÊ\begin{tabular}{|c|c|c|c|} \hline
ÊÊÊÊÊÊÊÊ$\beta_e$ & N &$\alpha$ & $\Lambda_b$ \\ \hline
ÊÊÊÊÊÊÊÊ0.01 - 0.4Ê & 12,507 & 0.51Ê & 0.40ÊÊÊÊÊÊ \\ 
ÊÊÊÊÊÊÊÊ0.4 - 1.5ÊÊ & 50,056 & 0.31Ê & 0.27ÊÊÊÊÊÊÊÊÊ \\ 
ÊÊÊÊÊÊÊÊ1.5 - 5.0ÊÊ & 80,560 & 0.20Ê & 0.18ÊÊÊÊÊÊÊÊÊ \\ 
ÊÊÊÊÊÊÊÊ5 - 100ÊÊÊÊ & 12,059 & 0.13 & -ÊÊÊÊÊÊÊÊ \\ \hline
\end{tabular}
\caption{The $\beta_e$ dependence of  $T_e(r) \sim r^{-\alpha}$ and the SH-collisionless transition scale $\Lambda_b L_T$.  N is the number of 
measurements in each $\beta_e$ interval.  Note that the limiting heat flux in the collisionless regime is proportional to the breakpoint 
$q_\parallel \simeq 1.07 ~\Lambda_b ~q_0$.  Also note that $2/7 \approx 0.286$.}
\end{center}
\end{table}

\section{Conclusions}

In this manuscript, we show that the transition from collisional (Spitzer-H\"{a}rm) to collisionless 
 electron heat conduction in the solar wind occurs at a mean free path of about $\lambda_{fp} \simeq 0.28 ~L_T$, where 
$L_T$ is the electron temperature gradient scale.  In the collisionless regime, the heat flux is limited to $q_\parallel \sim 0.3 ~q_0$, 
where $q_0$ is the free-streaming value.  
Some previous analyses suggested departures from the SH value \citep{Feldman++75, Pilipp++87}.  
This could be attributed to the choice of data intervals in those analyses, none of 
which were very statistical.  In fact, a preliminary statistical analysis of the Helios electro heat flux measurements shows 
similar results to what we present here (K. Horaites, private communication).


In our analysis, we have used the 'core' electron density and temperature to compute $q_0$ and $\lambda_{fp}$.  Theoretical work 
on heat conduction on steep temperature gradients has shown that a self-consistent flux limitation arises as SH theory begins to 
break down (for $f_1 \sim f_0$) and that this corresponds to order $\lambda_{fp}/L_T \sim 0.1$ \citep{Shvarts++81}, similar to 
our results.  The growing departure of the measurements from the SH curve in our Figure 2 is consistent 
with this effect (which can be seen by multiplying Figure 2 of \citep{Shvarts++81} by $\lambda_{fp}/L_T$).  
Since $f \sim n/T^{3/2}$, the constraint $f_1 \leq f_0$ is well-satisfied for solar wind parameters 
($n_h \sim 1/20 ~n_c$ and $T_h \sim 7-10 ~T_c$) for velocities less than about $v \lesssim 2.6 ~v_{th,c}$, therefore the core population does represent the collisional 
physics.  
Similar results were obtained by Smith et al. (2012), who solved the electron kinetic equation with a linearized Fokker-Planck operator in a 
fixed ion profile and found limiting heat flux values comparable to ours and attribute it to skewness in the distribution function at speeds of 
$v \geq 3 ~v_{th}$.
A simple model of a suprathermal tail escaping from an interplanetary 
electric field \citep{Hollweg74} predicts a collisionless heatflux $q \sim 3/2 ~n_e ~k_\mathrm{B} T_e ~v_{sw}$, which is also 
approximately consistent with our results, since $v_{sw} \sim 1/3 ~v_e$.
If we repeat our analysis with the full electron temperature $T_{all} = (n_c T_c + n_h T_h + n_s T_s)/(n_c + n_h + n_s)$, 
we find a similar breakpoint, but somewhat faster departure from SH.  

Our measurements show that the solar wind heat flux is well-described by the collisional Spitzer-H\"{a}rm value until the mean 
free path is approximately one third of the temperature gradient scale $\lambda_{fp} \sim 1/3 ~L_T$ and for larger $\lambda_{fp}$ is proportional to the 
saturation value $q \sim 1/3 ~q_0$.  While the limiting mechanism is not yet understood, these results should be useful for solar wind and coronal 
modeling efforts.  Our analysis also suggests that the $\beta_e$-dependence of whistler heat flux instability \citep{Gary++94} is 
inconsistent with the data, however a magnetosonic instability may be consistent in the collisionless regime.  We also 
infer a $\beta_e$-dependance to the temperature profile index $\alpha$, which may indicate an additional energy transport 
process, or be a proxy for another plasma parameter (e.g. Mach number, collisionality, etc.).

The observation that the electron heat flux remains 'classical' to Knudsen numbers as large as $K_T \sim 0.3$ may have 
implications for solar wind models which transition from fluid to exospheric domains  \citep{Echim++11} and for overall 
heat transport in the corona.  If the electron temperature gradient is only a weak function of radial distance (in the free solar wind), 
then $\lambda_{fp}$ will be smaller in the inner heliosphere, and it may be that most of the solar wind lies in the SH regime there.

Electron thermal conduction, in both the collisional Spitzer-H\"{a}rm and collisionless limits,   is energetically and dynamically important in other low-collisionality astrophysical plasmas, including, e.g., the hot intracluster medium in galaxy clusters \citep{Bertschinger++86}, the hot interstellar medium in galaxies (Cowie \& McKee 1977), and some accretion disks around neutron stars and black holes \citep{Sharma++06}.   These astrophysical plasmas are probably characterized by $\beta_e \sim 1-10$, not too dissimilar from a large fraction of the epochs of in situ solar wind data used here (Table 1).   Our results suggest that parallel thermal conduction is likely to be comparable to the Spitzer-H\"{a}rm and/or saturated values in these systems.   In more detail, the results in Fig. 3 and Table 1 could be used for modeling parallel thermal conduction in other astrophysical contexts.   Of course,  the global magnetic field geometry determines in part how the parallel thermal conductivity translates into large-scale redistribution of heat (e.g., Chandran \& Maron 2004), and the field geometry is significantly more uncertain in these other astrophysical contexts.

\acknowledgments

This work was supported in part by NSF grant AGS 0962726 (SHINE) and NASA HTP grant NNX11AJ37G.

\end{document}